\documentstyle[12pt]{article}
\begin{document}
\begin{center}
{\Large Antisymmetry and channel coupling contributions
to the absorption for p + $\alpha$/d + $^3$He}
\vspace{20mm}

{\large S.G. Cooper}%
\vskip1.0 cm%
{Physics Department, The Open University,\\ Milton Keynes
 MK7 6AA, UK}
\end{center}

{\bf abstract:} To understand recently established empirical p +
$\alpha$ potentials, RGM calculations followed by inversion are made
to study contributions of the d + $^3$He reaction channels and
deuteron distortion effects to the p + $\alpha$ potential. An
equivalent study of the d + $^3$He potential is also presented.  The
contributions of exchange non-locality to the absorption are simulated
by including an phenomenological imaginary potential in the RGM. These
effects alone strongly influence the shape of the imaginary potentials
for both p + $\alpha$ and d + $^3$He.  The potentials local-equivalent
to the fully antisymmetrised-coupled channels calculations have a
significant parity-dependence in both real and imaginary components,
which for p + $\alpha$ is qualitatively similar to that found
empirically.  The effects on the potentials of the further inclusion
of deuteron distortion are also presented.  The inclusion of a
spin-orbit term in the RGM, adds additional terms to the phase-equivalent
potential, most notably the comparatively large imaginary spin-orbit
term  found empirically.


\vfill
\hfill \today

\pagebreak



\setlength{\parindent}{0.3 in}

\section{Introduction}

In successive studies, application of Iterative Perturbative (IP)
inversion to empirical p + $\alpha$ $S$-matrices has established a
local potential, which is both parity and energy dependent, covering a
wide energy range from subthreshold energies up to $\sim 70$
MeV,~\cite{pa-emp,pa-edept}.  For energies above the inelastic
threshold, parity dependence is also necessary in the imaginary
components to reproduce the empirical behaviour of
$|S(l,j)|$. Furthermore, a recent energy dependent inversion for p +
$^{16}$O found an imaginary parity dependent component essential to
reproduce satisfactorily a comprehensive set of elastic cross-section
and analysing power data, \cite{p-ox16}. Parity dependence in the real
N + $\alpha$ potential has been widely studied and is understood to
simulate particular exchange effects, notably heavy-particle pickup,
arising through the antisymmetry, \cite{na-exch}.  At subthreshold
energies, a close correspondence has been established between the real
potential local equivalent to Resonating Group Model (RGM) phase
shifts and parity dependent potentials determined by inversion from
empirical phase shifts, \cite{pa-rgm,n-nucl}.  It is now of interest
to know how far fully antisymmetrised coupled channels calculations
can explain the form of imaginary parity dependent components found
empirically and this is a major aim of the present study.  In
including coupling to the d + $^3$He channel, the d + $^3$He
$S$-matrices are also calculated and a parallel study of the d +
$^3$He potential is made possible.

This study is based on RGM calculations followed by IP $S(l)
\rightarrow V(r)$ inversion. The 5 nucleon system has been subject to
many microscopic investigations (for example, \cite{Reichstein}-
\cite{pa-res}).  At subthreshold
energies the single channel approximation provides qualitatively
reasonable results, while at higher energies, increasing numbers of
reaction channels are required.  Inevitably, such calculations will
always necessitate some unknowns or approximation, for example in the
cluster wavefunction basis or the nucleon nucleon potential, so that
precise fit to experimental results remain difficult to achieve. Even
the most elaborate of recent RGM calculations, \cite{KKT-95}, with up
to 25 cluster configurations, cannot reproduce the p + $\alpha$
reaction cross-sections.  No attempt is then made here to reproduce the
empirical potentials precisely and the current aim is to establish
general features in the potential which arise through the non-locality
of the antisymmetry and channel coupling. The only coupling
included is between the p + $\alpha$ and d + $^3$He channels and with
the excited deuteron pseudo-states, which represent s-wave deuteron
breakup following Ref.~\cite{RGM-5n-d}. These RGM calculations do not
include a tensor force, which is expected to be of greatest importance
at energies close to the d + $^3$He threshold,
\cite{pa-res}, The current calculations are made  at energies well 
above the d + $^3$He threshold.

At the simplest level, absorption effects have been simulated by
introducing a phenomenological imaginary potential into the single
channel RGM (see for example Ref.~\cite{RGM-dhe}).  The absorption in
the resulting $S(l)$ will be modified by the non-locality due to
exchange contributions, and the imaginary potential resulting
from subsequent $S(l) \rightarrow V(r)$ inversion may differ
substantially from the phenomenological imaginary potential.  As will be
shown, this non-locality can even lead to the imaginary potential
becoming emissive at the origin.  These induced effects depend on the
system studied and, in Sect.~\ref{sect-impots}, results are compared
for p + $\alpha$ and d + $^3$He scattering.

The outline of the paper is as follows: Section~\ref{rgm-defs} outlines
details of the RGM plus inversion procedure  and establishes the notation
used in the rest of the paper.  The results of using an
imaginary potential in single channel RGM calculations are presented
in Section~\ref{sect-impots}. The contributions of the deuteron
distortion channels alone to the d + $^3$He potential are discussed in
Section~\ref{d-dist} and the results of including reaction channels
into the RGM for both p + $\alpha$ and d + $^3$He are described in
Section~\ref{react-ch}. The paper ends with a short summary section.

\section{RGM plus inversion calculations}
\subsection{The RGM calculations}
\label{rgm-defs}

The RGM calculations use a modified form of the codes of Bl\"{u}ge
{\em et al}, \cite{RGM-code}. The original code permits RGM
calculations in both single channel mode and with RGM channel coupling
between the p + $\alpha$ and d + $^3$He channels.  Deuteron distortion
effects may additionally be included using the pseudo-state method of
Kanada {\em et al}, \cite{ad-dist}.  The code of Bl\"{u}ge {\em et
al\/} has been rewritten in Fortran 90 and adapted to incorporate an
alternative choice of nucleon-nucleon potential, an imaginary
phenomenological potential and a larger basis for the deuteron
wavefunction, which consequently permits the inclusion of more
pseudo-states.  The code cannot incorporate a tensor interaction in
the nucleon-nucleon force, so that the reaction channel coupling is
allowed only between p + $\alpha$ and the d + $^3$He configuration
with channel spin $s=1/2$, i.e.  there can be no coupling to the d +
$^3$He channel with $s=3/2$ . However, calculations without reaction
channel coupling are presented for the d + $^3$He spin $3/2$ channel,
to allow a direct comparison of the potentials for the two channel spins.

In the RGM calculations presented here, the internal motions of the
$\alpha$ and $^3$He are described by single gaussian form-factors of
widths, 0.606 and 0.367 respectively.  These widths give reasonable
estimates for the root mean square matter radii of the respective
nuclei, but the corresponding energy expectation values are -24.72 and
-3.51 MeV respectively for the $\alpha$ and $^3$He, a little greater
than the experimental values. These ground state energies put the d +
$^3$He threshold at 19.0 MeV, again slightly higher than the
experimental value of 18.36 MeV, \cite{A-selove}.  However, since a
precise agreement with empirical results is not the aim of this work,
this disagreement should not be very significant.
The deuteron wavefunction is represented by an 8 gaussian function,
\cite{kuk-Ryz}, which was constructed to have several low energy deuteron
excitation modes, \cite{kuk-priv}. The corresponding ground and
excited deuteron energies are -2.202, -0.036, 0.2008, 0.888 and 3.94
MeV, and 4 orthogonal pseudo-states may be used in calculations
to represent deuteron distortion.

All the RGM calculations are based on the nucleon-nucleon potential of
Thompson {\em et al\/}, \cite{minn-pot}, with the exchange mixture
parameter $u=0.95$.  For simplicity, most of the results presented in
the following sections are obtained without a spin-orbit interaction.
Where specifically stated, calculations include the nucleon-nucleon
spin-orbit force of Reichstein and Tang, \cite{Reichstein}.

The following notation is used to simplify the discussion in the rest
of this paper: Single channel calculations, leading to real phase shifts
and potentials are denoted ``SC''. RGM calculations which include
reaction channel coupling only, without distortion effects, are
denoted ``CC''.  The d + $^3$He calculations with coupling to deuteron
pseudo-states only are labelled ``DC'' and the complete calculations,
in which both the p + $\alpha$ and d + $^3$He channels are coupled to
the four deuteron pseudo-states are denoted ``6C''.

An imaginary central potential can be included in the RGM calculations
as a direct addition to the real RGM direct potential, following the
procedure described by Chwieroth {\em et al}, \cite{RGM-dhe}.  In the
present work, the imaginary phenomenological potential has a gaussian
form, i.e.,
\begin{displaymath}
  W(r) = W_0 \exp \left\{- \left(\frac{r-R_0}{a}\right)^2\right\}.
\end{displaymath}
and, except where otherwise specified, $W_0 = 2$ MeV, $R_0 = 1$ fm
and $a = 2$ fm. These parameters define a p + $\alpha$ imaginary
potential which is reasonably close in magnitude and shape to that
found empirically.  The aim of the present calculations is only to
investigate effects on the resulting absorption due to
antisymmetry and no attempt is made to compare RGM calculations
with experimental cross-sections.  In the following sections, the
above imaginary phenomenological potential is included in RGM single
channel calculations only, for both p + $\alpha$ and d + $^3$He,
$s=1/2$ and $s=3/2$ cases, and these calculations are denoted ``IM''.

\subsection{The RGM $S(l)$}

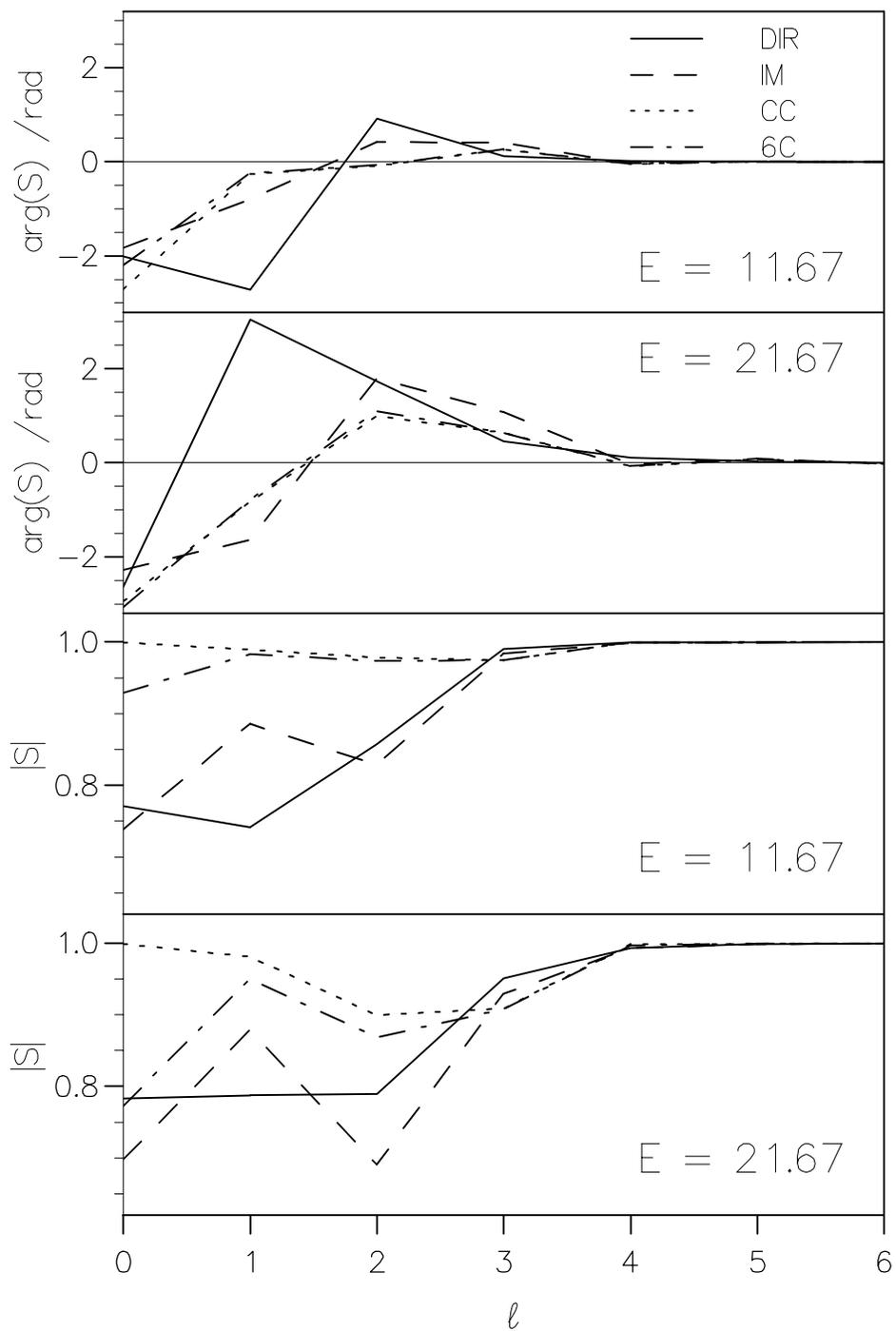
\begin{figure}
\caption{For d + $^3$He, values of $|S_l|$ and $\arg(S_l)$ calculated
at energies of 11.67 and 21.67 MeV for the following cases (defined in
Sect.~\ref{rgm-defs}), DIR (full lines), IM (dashed
lines),  CC (dotted lines) and 6C (dash-dot lines).}
\label{fig-sme-dh}
\end{figure}

Values of $|S_l|$ and $\arg(S_l)$ for selected calculations are shown
in Figs.~\ref{fig-sme-pa} and \ref{fig-sme-dh}, for p + $\alpha$ and d
+ $^3$He respectively. In each case $S(l)$ is displayed for two
energies, the energy of the $S(l) \rightarrow V(r)$ inversion and 10
MeV lower in energy to provide an indication of the energy dependence.
For p + $\alpha$, inversions are presented at 40 MeV (in the
laboratory frame), well above the reaction threshold at 23 MeV  This energy
corresponds to a deuteron laboratory energy of 21.67 MeV, given
the RGM form-factors used here, and the
inversions for d + $^3$He are mostly presented for this energy.

An additional case, labelled ``DIR'', is included in Figs.~\ref{fig-sme-pa} and
\ref{fig-sme-dh}. The DIR potential comprises the sum of the
phenomenological imaginary potential and the real RGM direct nuclear
potential (obtained by the double folding of the N-N
potential with the appropriate wavefunctions, as defined in
Ref.~\cite{RGM-dhe}). The DIR $S$-matrix is obtained from the direct
insertion of the DIR potential into the schr\"odinger equation,
i.e. without antisymmetry.  The values of $\arg(S)$ for this DIR
case are almost identical to the values obtained by a similar direct
calculation without the imaginary potential. The large differences
between $S(l)$ for the DIR and the SC cases is well known and has been
long understood to arise predominantly due to what is commonly known
as the ``knock-on exchange'' term, \cite{LeMere}.

Figs.~\ref{fig-sme-pa} and \ref{fig-sme-dh} are discussed in greater
detail in the following sections.  One obvious feature is the
relatively low absorption in the p + $\alpha$ case, for which
$|S(l=2)| > 0.85$ at $E=40$ MeV. This $|S|$ is considerably greater
than the empirical values of $\sim 0.65$ and $\sim 0.76$ for $j = 3/2$
and $j = 5/2$ respectively, \cite{plattner,burzynski}.  This
deficiency probably arises due to the absence of tensor coupling which
establishes the resonance at the reaction threshold and also has a
large effect on $|S|$ for the d$_{3/2}$ partial wave immediately above
the threshold \cite{burzynski}.

\subsection{$S$-matrix to potential inversion}

All the $S$-matrix to potential inversions use the
Iterative Perturbative code IMAGO, \cite{imago}, and the procedure is
described in many previous publications (see Ref.~\cite{pa-edept}
and refs therein).  The success of the inversion is measured by a
phase shift distance, $\sigma^2$, between the target (RGM) $S$-matrix,
$S^{\rm tar}$, and the $S$-matrix $S^{\rm inv}$ corresponding to the
potential established by inversion, i.e.
\begin{equation}
 \sigma^2 = \sum_{\kappa}  |S_{\kappa}^{\rm tar} -
S_{\kappa}^{\rm inv}|^2 \label{sig}
\end{equation}
and where the index $\kappa$ represents the $l$, $j$ and energy values
of a specified included partial wave.  The inversion can then be made
for an $S$-matrix at a single energy or for $S$-matrices for a range
of energies. Parity dependent potentials are established in the form,
$V(r) = V_1(r) + (-1)^l V_2(r)$.  Since few partial waves contribute
to the inversion for both p + $\alpha$ and d + $^3$He, only a small
inversion basis of 3 -- 4 basis functions can be applied for each
potential component. Use of a gaussian basis generally leads
to the smoothest potentials, but there is necessarily a strong
sensitivity to the choice of basis. The resulting uncertainties are
most noticeable near $r=0$ where there is an ambiguity between the
$V_1$ and $V_2$ terms. In all cases the smoothest potentials obtained
are presented in the following sections.

The inclusion of spin-orbit coupling in the d + $^3$He channel spin
$s=3/2$ calculations, presents a new challenge for the iterative
perturbative inversion procedure in the form of spin-3/2
inversion. The inversion method is easily adapted if tensor terms are
omitted. The main challenge remaining is then to reproduce phase
shifts for more $l,j$ values than the cases with lower spin, but using
only central and spin-orbit potentials.

\section{Effects due to a phenomenological imaginary potential}
\label{sect-impots}

\subsection{p + $\alpha$ scattering}
\label{pa-impots}

The comparison of  the DIR and  IM $S(l)$ for p +
$\alpha$ in Fig~\ref{fig-sme-pa}, shows the contributions of
antisymmetry to both $\arg S$ and $|S|$, which, in this case,
effectively relate to the real and imaginary local potentials
respectively.  The introduction of the exchange effects produces a
strong parity-dependent effect in $\arg S(l)$, still conspicuous at 40
MeV lab energy, which is little changed by this simple introduction of
absorption. However, for the s and p waves, $|S|$ obtained for the IM
case is significantly closer to unity than $|S|$ for the DIR case.
This behaviour is characteristic of the Perey effect, \cite{edep-na},
which arises due to knock-on exchange.

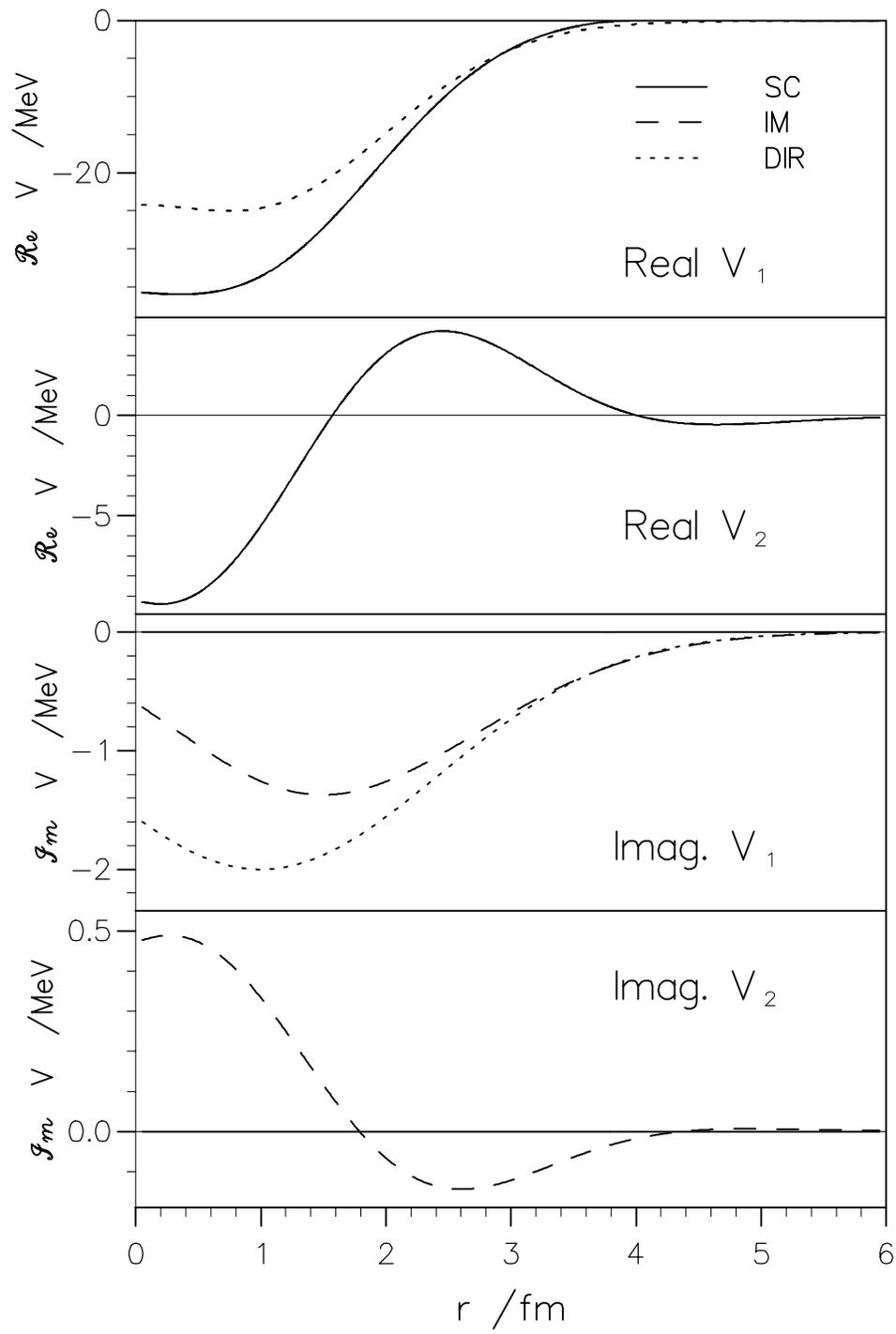
\begin{figure}
\caption{For p + $\alpha$ at a laboratory energy of 40 MeV,
the real and imaginary, $V_1$ and $V_2$ components obtained by inversion of
the SC $S(l)$ (full lines) and the IM $S(l)$ (dashed lines), compared
with the parity independent DIR potential (dotted lines).}
\label{fig-pa-im}
\end{figure}

Application of IP Inversion to the SC and IM $S(l)$ at 40 MeV 
yields parity dependent potentials, which are
complex for the IM case. These potentials are displayed in
Fig,~\ref{fig-pa-im}, together with the DIR potential. The values of
$\arg S$ are practically identical for the SC and IM cases, so that
inversion leads to real potentials for these two cases which are
graphically indistinguishable.  These real potentials are very similar
to potentials previously determined from single channel RGM phase
shifts for this energy range, \cite{pa-rgm,n-nucl}.  Near the nuclear
centre the potential is deeper than the direct potential, reflecting
the increase in $\arg S$ for $l <2$ induced by the exchange
effects. For $l=2$, $\arg S$ is decreased when antisymmetry is
included.

The difference in magnitude, for $ r < 2$ fm, between the imaginary
potential obtained by inversion and the DIR imaginary component
follows in consequence of the behaviour of $|S|$ for $l < 2$.  If a
more surface peaked imaginary potential is used in the RGM, the
potential local-equivalent to the resulting phase shifts may become
emissive at $r=0$.  An imaginary $V_2$ term must be included in the
inversion to reproduce the phase shifts accurately, but this term is
clearly small, that is, much less than the corresponding empirical
term. It then appears that the heavy-particle pickup, which causes the
real parity-dependence, is of lesser importance to this imaginary
component.

The addition of a nucleon-nucleon spin-orbit term into the RGM does
not produce any significant changes to the above results.  The
exchange non-locality also has very little effect on the spin-orbit
terms of the resulting local potential. The real $V_1$ spin-orbit term
obtained by inversion is almost identical to the RGM direct spin-orbit
term, \cite{Reichstein}.  The inclusion of parity-dependent and
imaginary spin-orbit terms in the inversion does decrease $\sigma$,
but the resulting potentials are small, i.e. are $< 1$ MeV for the
real $V_2$ term and $ < 0.2$ MeV for the imaginary components.

\subsection{d + $^3$He scattering, $s=1/2$}
\label{scim-dh-s0p5}

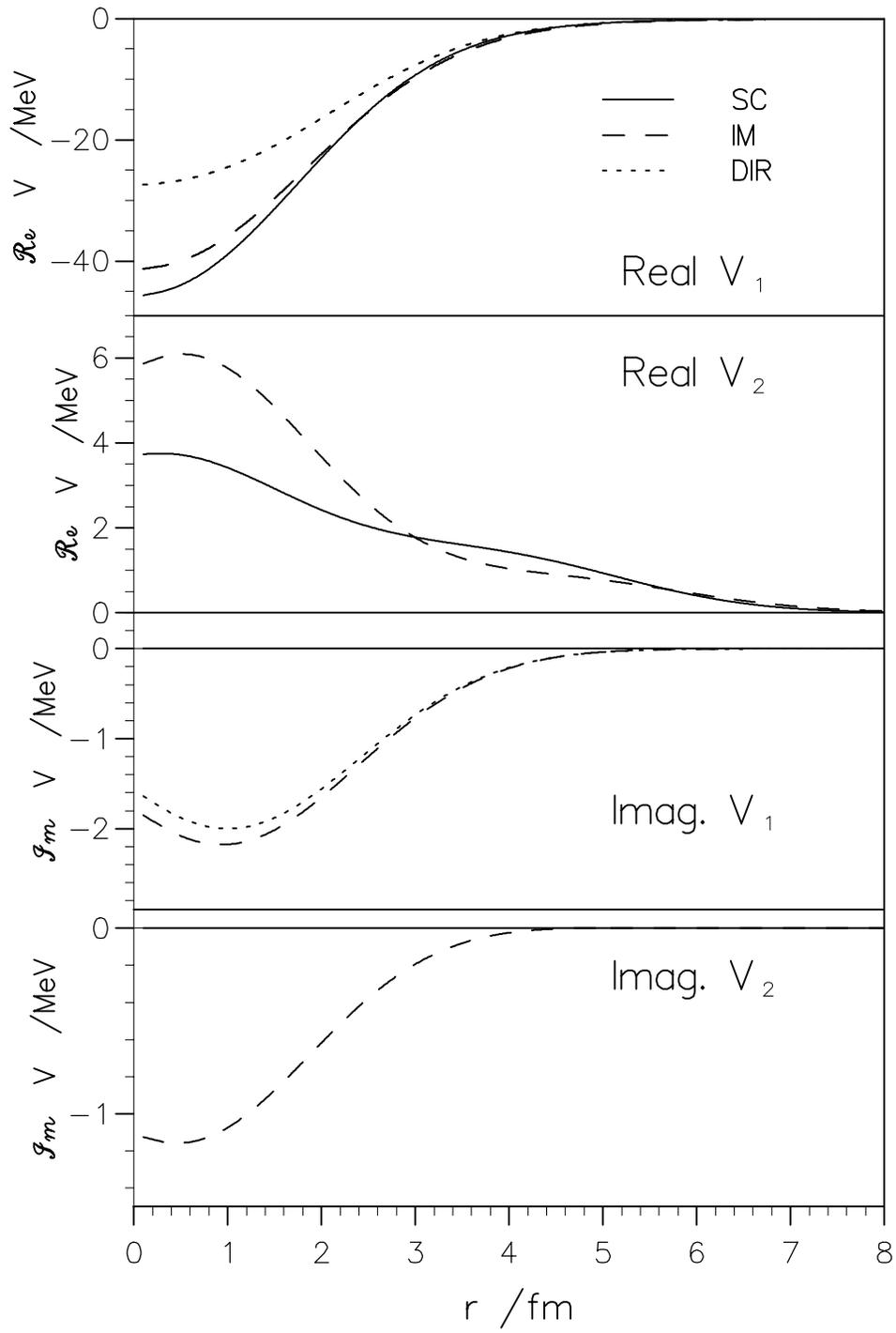
\begin{figure}
\caption{For d + $^3$He at a deuteron laboratory energy of 
21.67 MeV and channel spin $s=1/2$, the real and imaginary, $V_1$ and
$V_2$ components obtained by inversion of the SC $S(l)$ (full lines) and
the IM $S(l)$ (dashed lines), compared with the parity independent DIR
potential  (dotted lines). }
\label{fig-dh-im}
\end{figure}

Figs.~\ref{fig-sme-pa} and \ref{fig-sme-dh} provide an indication of
the differences between the effects of exchange on the $S$-matrices
for d + $^3$He and for p + $\alpha$.  The local potentials determined
by inversion for d + $^3$He, with channel spin $1/2$, at $21.67$ MeV
are displayed in Fig.~\ref{fig-dh-im} for the DIR, SC and IM cases.  

Fig.~\ref{fig-dh-im} shows a large increase in the depth of the real
central $V_1$ component for the SC and IM cases compared with the
corresponding DIR potential. This decrease  presumably arises due to
``knockon-exchange'', as in the p + $\alpha$ case.  However, the $V_2$
component resulting from inversion of the SC and IM cases is entirely
positive for d + $^3$He.  The behaviour of $\arg S$ for the SC case is
again almost identical to that the SC case, except for $l=0$.  This
small difference in one $\arg S$ value leads to large differences
between the SC and IM cases in the real potentials determined by
inversion, as shown in Fig.~\ref{fig-dh-im}.

The $|S|$ for the IM case significantly differ from the $|S|$ for the
DIR case for $l < 4$, but not in the form of a clear increase in $|S|$
as seen for p + $\alpha$. This suggests that $|S|$ may be
significantly influenced by more than just the knock-on exchange
contribution.  The resulting $V_1$ imaginary term obtained by inversion is
very close to the DIR potential and an imaginary $V_2$ component is
induced which is about half the magnitude of the $V_1$ component at
$r=1$ fm.  If the parity-dependence arises predominantly due to the
heavy-particle pick-up then this exchange effect must be considerably
greater for d + $^3$He than for p + $\alpha$.

The extended radial structure of the deuteron wavefunction contributes
significantly to the differences found between the p + $\alpha$ and d
+ $^3$He cases.  The two systems show a very different behaviour for
the divergence of the probability current, $\nabla
\cdot{\mathbf j(r)}$ as conventionally defined (and calculated as described
in Ref.~\cite{n-wavefns}).  When evaluated for the full $s=1/2$ d +
$^3$He wavefunction, $\nabla \cdot{\mathbf j(r)}$ is not only
substantial for a much wider radial range than the corresponding
calculation for p + $\alpha$, but has also at least twice the
magnitude at the maximum value.

A very different pattern of potential features is found if the SC and
IM calculations $S(l)$ are based on a simpler deuteron wavefunction
consisting of one gaussian alone (with a width = 0.4124 fm). For this
one gaussian case the behaviour of $S(l)$ is qualitatively similar to
the values of the full eight gaussian wavefunction, with small
differences most noticeable for $l=0$. The real potentials determined
for these new SC and IM cases show far more similarities, in shape and
magnitude, than the corresponding potentials shown in
Fig.~\ref{fig-dh-im}. Furthermore, the imaginary components obtained
by inversion for the new one gaussian IM case also have a different
behaviour to that displayed in Fig.~\ref{fig-dh-im}. The new imaginary
$V_1$ component now has a reduction in the magnitude near the nuclear
centre compared to the DIR case, similar to the corresponding
potential found for p + $\alpha$. The corresponding $V_2$ component
has a magnitude of only 0.25 MeV, much less than that obtained using
the full eight gaussian wave function. If the imaginary parity
dependence arises predominantly from heavy-particle pickup, as for the
real parity-dependent term, the probability of this exchange is
clearly considerably increased by the use of a deuteron wavefunction
with a larger radial extent.

The changes in $S(l)$ following the introduction of the
phenomenological imaginary term, and consequently in the potentials
determined by inversion, are energy dependent. This energy dependence
is introduced entirely by the exchange non-locality, since both the
nucleon-nucleon potential inserted into the RGM and the
phenomenological imaginary term are energy independent. However, a
quantitative assessment of the introduced energy dependence is
difficult.  There is a spurious resonance in the SC case for $l=0$ at
about 32 MeV, as found by Chwieroth {\em et al}, \cite{RGM-5n-im},
which is effectively removed by including absorption.  Energy
dependent inversion, using a linear energy dependence for the real
components, leads to an accurate reproduction of the IM $\arg S$ for
an energy range of $\sim 10$ MeV, centred at 40 MeV.  The resulting
real components at 40 MeV are very close to the corresponding
potentials shown in Fig.~\ref{fig-dh-im}.  However while the imaginary
potential input into the RGM has no energy dependence, the resulting
absorption shows a strong variation with energy for $l
\leq 2$, so that $|S|$ cannot be reproduced by inversion using simple
linear dependence on energy in the imaginary components.  Potentials
obtained by applying fixed energy inversion at a series of energies
show a decrease with energy in the magnitude of the imaginary $V_2$
term in the surface region, while both the imaginary $V_1$ and $V_2$
terms increase in depth at the nuclear centre as the energy increases.

The inclusion of a simple spin-orbit component in the nucleon-nucleon
potential inserted into the RGM calculations has little effect on the
results found above. The direct real spin-orbit component obtained
from the RGM is much smaller for d + $^3$He than for p + $\alpha$,
i.e. $\sim 1$ MeV at $r=1$ fm compared to $\sim 12$ MeV for p +
$\alpha$. Both with and without absorption, the exchange contributions
induce a small, but positive, parity dependent component for d +
$^3$He of $\sim 0.25$ MeV at $r=1$ fm, proportionally much larger than
for p + $\alpha$. Again this suggests a much stronger influence of the
exchange terms, particularly the heavy-particle pick-up term, for d +
$^3$He than for p + $\alpha$.  However, in the IM cases there is no
necessity for any significantly non-zero imaginary spin-orbit terms in
the local potential.

\subsection{d + $^3$He scattering, $s=3/2$}
\label{scim-dh-s1p5}

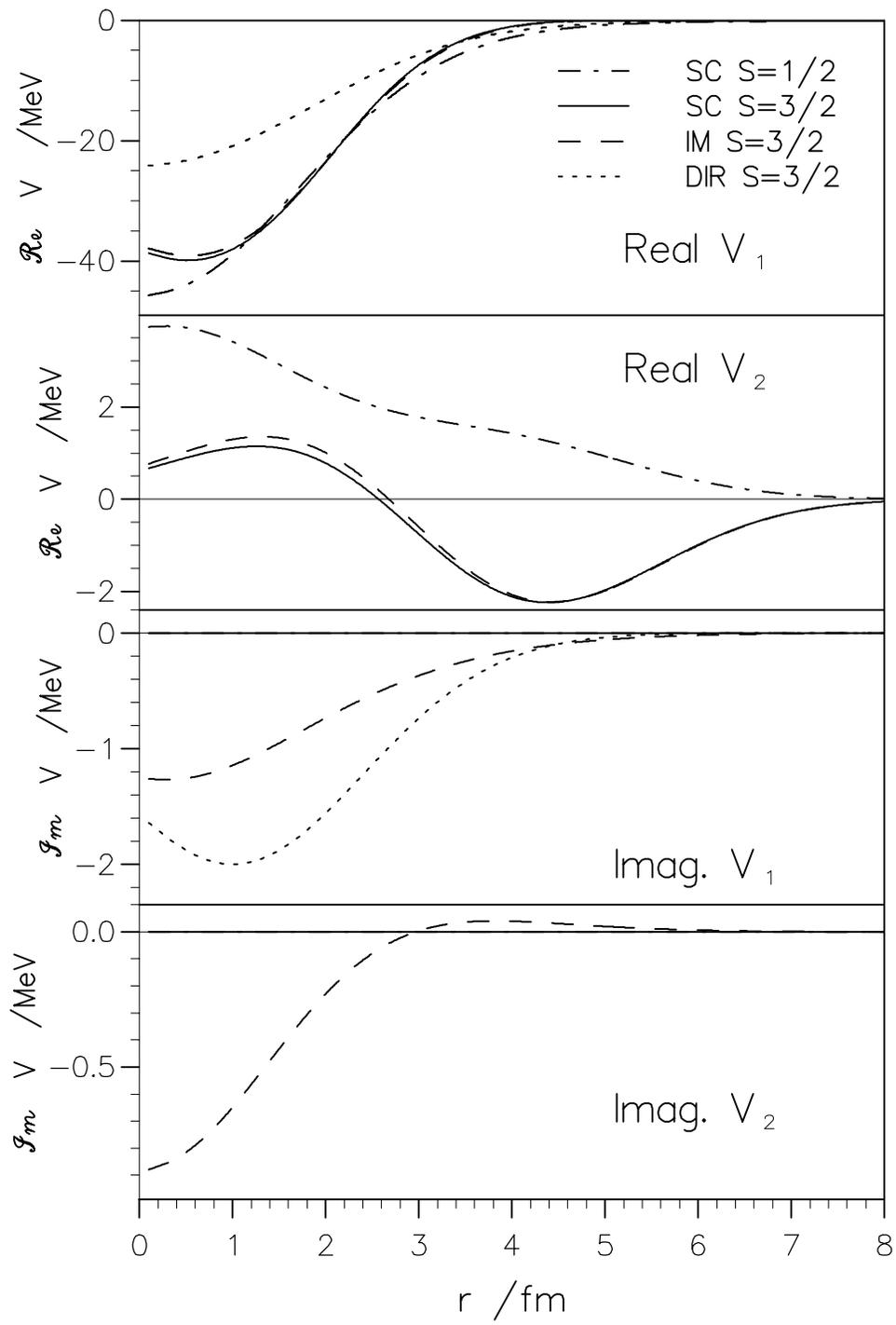
\begin{figure}
\caption{For d + $^3$He at a deuteron laboratory energy of 
21.67 MeV and channel spin $s=3/2$, the real and imaginary, $V_1$ and
$V_2$ components  obtained by inversion of the SC $S(l)$ (full lines) and
the IM $S(l)$ (dashed lines), compared with the parity independent DIR
potential  (dotted lines) and the
potential obtained by inversion of the SC $S(l)$ for 
$s=1/2$ (dash-dot lines).}
\label{fig-dh-s1p5}
\end{figure}

The strong dependence of the d + $^3$He phase shifts on channel spin
reported by Chwieroth {\em et al}, \cite{RGM-dhe}, is found here in
both the phase shifts and local-equivalent potentials, with and
without absorption. Chwieroth {\em et al} attribute these differences
to effects of the Pauli exclusion principle on the two channel spins.
As in the $s=1/2$ case, significant changes are introduced, in both
the RGM $S(l)$ and the resultant local equivalent potentials, when
exchange is included in the $s=3/2$ d + $^3$He RGM calculations.  The
corresponding potentials obtained by inversion for the SC, IM and DIR
calculations are displayed in Fig.~\ref{fig-dh-s1p5}, for a deuteron
laboratory energy of 21.67 MeV.  Fig.~\ref{fig-dh-s1p5} also includes
the SC potential determined for d + $^3$He with channel spin $1/2$.
The real $V_1$ components for the two channel spins in the SC case are
qualitatively very similar. However, while the differences for $r<1$
fm may be due to ambiguities in the inversion, at larger radii the
real $V_1$ component is clearly smaller in magnitude for $s=3/2$ than
for $s=1/2$.  The real $V_2$ component shows a more significant
difference between the two channel spins, since these components are
opposite in sign for $r > 2.5$ fm. This $V_2$ is therefore very
different from the p + $\alpha$ $V_2$ component, as predicted by
Chwieroth {\em et al}.

The values of $\arg(S)$ for the $s=3/2$ IM case are almost identical
to the values for the SC case for all $l$ and consequently there is
little difference between the SC and IM real potentials. As in
previous cases discussed above, the IM $|S|$ again differs
considerably from the $|S|$ for the DIR case. The imaginary $V_1$
potential consequently decreases in magnitude for $r < 3$ fm, by
comparison with the DIR potential, characteristic of a Perey-like
effect.  Unlike the p + $\alpha$ case, the corresponding imaginary
$V_2$ term is relatively large for $r<2$ fm, and is similar to that
obtained for the equivalent potential component for $s=1/2$.
Therefore, although very different effects are obtained for the two
channel spins, in both cases the non-locality generated by the
exchange contributions must be much stronger for both values of $s$
than in the p + $\alpha$ case. 

The introduction of a spin-orbit term into the single channel RGM
produces a much larger inversion problem for $s=3/2$. For this SC case
a low inversion $\sigma$ is possible by establishing only central and
spin-orbit, $V_1$ and $V_2$ terms. The central components are then
very similar to the equivalent real components shown in
Fig.~\ref{fig-dh-s1p5}.  As found in the $s=1/2$ case, inversion of
the d + $^3$He $s=3/2$ $S(l)$ requires a spin-orbit $V_2$ term, which
is large and negative, i.e. $\sim -0.5$ MeV at $r=1$ fm. This term
will also arise due to the strong non-locality resulting from
exchange.

\section{Deuteron Distortion Effects on the d + $^3$He potential.}
\label{d-dist}

Breakup effects are particularly important in d + $^3$He due to the
high compressibility of the deuteron.  In a detailed study of
distortion effects in d + $^3$He for channel spin $s=3/2$, Kanada {\em
at al\/}, \cite{RGM-dist}, showed that the specific distortion effects
contribute much more strongly to the Pauli favoured states then to the
Pauli unfavoured states. These observations were deduced from the
phase-shifts, but such behaviour must also be reflected in the
potential, especially in the parity-dependence.

In this section the contributions to the d + $^3$He potential arising
only due to deuteron distortion are considered.  The distortion
contributions are calculated for both channel spins, although for
$s=1/2$ the $S(l)$ will not be physically realistic because the p +
$\alpha$ channel coupling is the more energetically favoured
configuration. The most accurate results from the RGM have been
obtained with up to 15 distortion channels, but a reasonable
approximation is found by including only the states corresponding to
the lowest 5 energies, \cite{RGM-5n-d}.  The most significant effect
of increasing the number of distortion channels is to reduce the
waviness in the energy dependence of $S(l)$, which arises from
unphysical dispersion-like resonances.  Calculations based on all 5
deuteron states possible with the present choice of deuteron
wavefunction (DC) do not have the spurious resonance in the s=1/2
channel for $l=0$ at $\sim 19$ MeV lab energy. Some smaller
and narrower resonance-like features are introduced by the
pseudo-state coupling.  While the $s=1/2$ channel shows a greater
energy variation in $|S|$, this variation is mostly of significance
outside the energy range of the $S(l) \rightarrow V(r)$
inversion. However, for the $s=3/2$ case, a resonance appears for
$l=0$ at $\sim 21.5$ MeV.

Inversion of the DC d + $^3$He RGM phase shifts is based on a wider
energy range than for the potentials presented in the preceding
sections to avoid obtaining unreasonable deductions due to the
spurious resonances. Specifically inversion is applied to $S(l)$ for
the range of deuteron laboratory energies, $16.6 < E_d < 26.7$.  By
including a linear energy dependence in the real potential terms, an
accurate reproduction of $\arg (S)$ is maintained. However, no simple energy
dependence is found for the imaginary components and the resulting
potentials provide only an average fit to $|S|$ over the included
energy range.

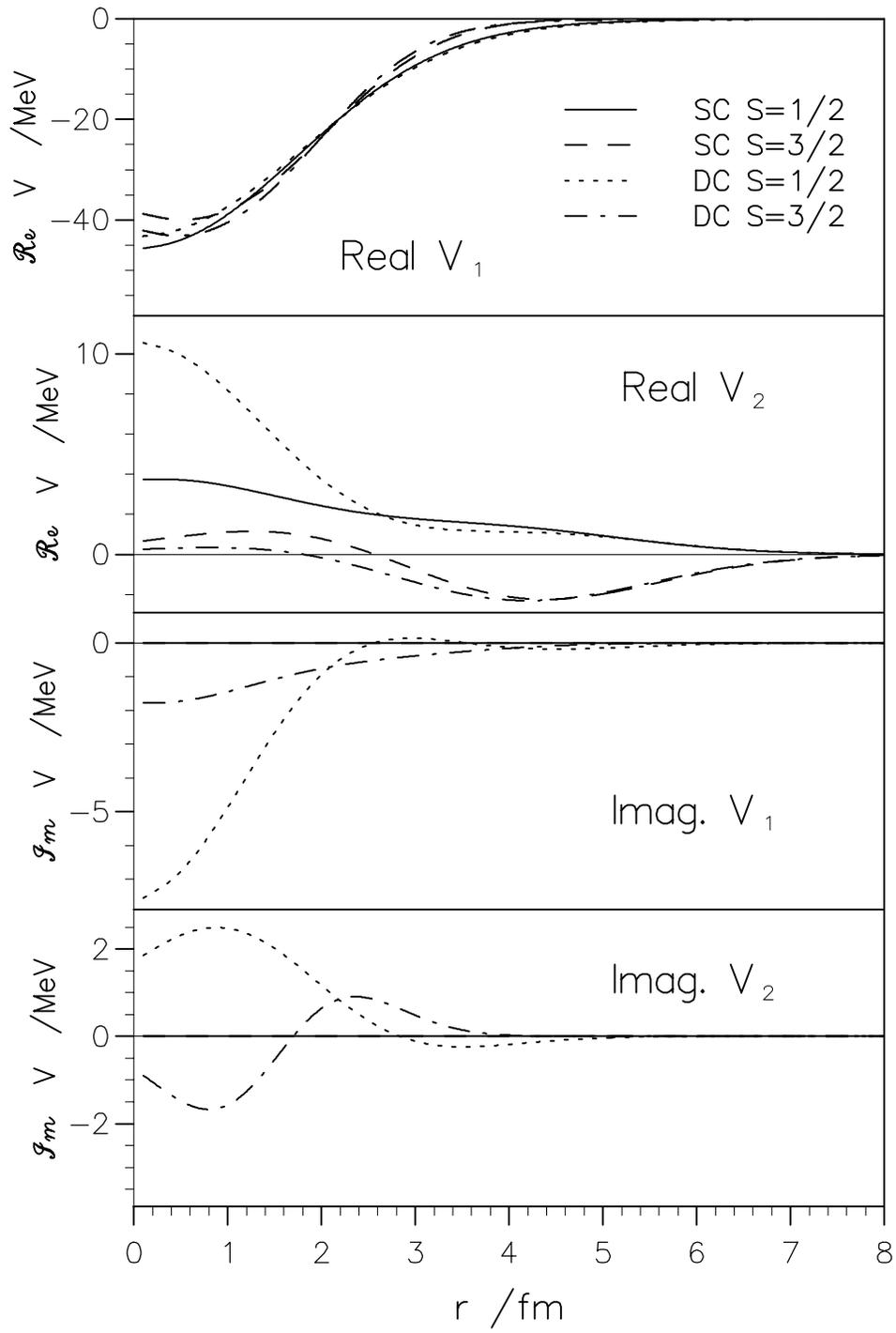
\begin{figure}
\caption{For d + $^3$He, the real and imaginary, $V_1$ and $V_2$ components 
obtained by inversion of (i) the SC $S(l)$ for a deuteron laboratory
energy $E_d = 21.67$ MeV, for both $s=1/2$ (full lines) and $s=3/2$
(dashed lines), and (ii) the DC $S(l)$ over the energy range $16.6 <
E_d < 26.7$ MeV for both $s=1/2$ (dotted lines) and $s=3/2$ (dash-dot
lines).}
\label{fig-dh-dist}
\end{figure}

The potentials, evaluated at an energy of $21.67$ MeV, for both
$s=1/2$ and $s=3/2$ are presented in Fig.~\ref{fig-dh-dist}, which
also shows the potentials determined for the SC cases.  The most
significant effects of distortion appear at small radii, unsurprisingly
as the distortion effects are strongest for low $l$,
\cite{RGM-dist}.  The ambiguities between the real $V_1$ and $V_2$
components makes clear deductions difficult, but for $s=1/2$ there is
a substantial increase in the $V_2$ component which has no parallel feature
for $s=3/2$. These changes are very similar to the changes induced in
the real potential by the introduction of the simple absorption term
(the IM case).  At large radii the distortion contributions show little
effect.

The magnitude of the imaginary terms for $s=3/2$ shown in
Fig.~\ref{fig-dh-dist} are ill-determined due to the resonance
dominating $|S|$ in the energy range considered.  However a strong
parity dependent component is required in fixed energy inversion at
any energy within this range, and parity-dependent terms are also
required for $s=1/2$. These imaginary $V_2$ components are similar in
magnitude but very different in shape for the two channel spins. The
two potentials also differ in sign at small radii, notably in the
radial region where both real $V_2$ components are positive.  Clearly
here the Pauli effects are strong, but these results also show the
contributions of deuteron distortion to be of significance for the
potentials for both values of $s$.

\section{Contributions from the reaction  channels}
\label{react-ch}

\subsection{p + $\alpha$}

The contributions of the coupling to the d + $^3$He channel on the p +
$\alpha$ RGM potential have previous been investigated at subthreshold
energies, \cite{pa-rgm}.  While the coupling contributes just small
changes in the potential magnitude, only when the d + $^3$He coupling
is included is the decrease with energy, found empirically for the
real components, obtained for the RGM potentials.  
At higher energies,
the reaction channel coupling contributions to $\arg(S)$ remain small,
as seen in Fig.~\ref{fig-sme-pa}, but above 30 MeV, the contributions
to the absorption become important for the d and f waves. For $l < 2$,
$|S|$ remains close to unity, similar to the behaviour in the
simpler IM case.  The introduction of distortion (6C case) in the d +
$^3$He channel introduces additional uncertainties in the phase shifts
due to the necessity to extend the radial integration out to very
large radii, \cite{KKT-95}.  This extra coupling does not, however, lead
to significant changes in the p + $\alpha$ $S(l)$.

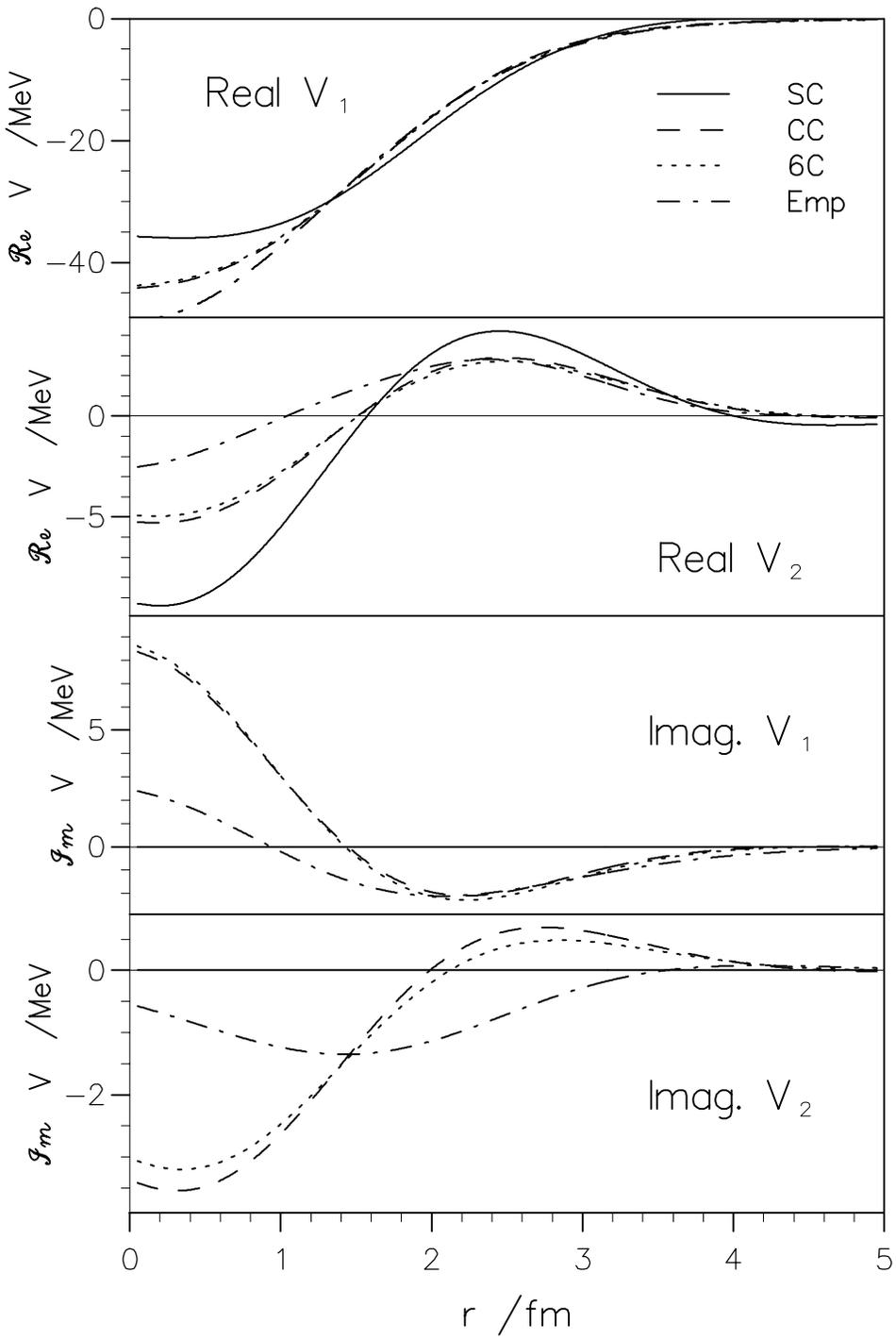
\begin{figure} 
\caption[For p + $\alpha$ at a laboratory energy of 40 MeV]
{For p + $\alpha$ at a laboratory energy of 40 MeV,
the real and imaginary, $V_1$ and $V_2$ components obtained by inversion of
the CC $S(l)$ (dashed lines) and 6C $S(l)$ (dotted lines) compared
with the equivalent components determined for the SC case (full lines)
and an empirical potential determined in Ref.~\cite{pa-edept}, (Burzynski (6))
labelled ``Emp''.}
\label{fig-pa-cc5c}
\end{figure}

The potentials obtained by inversion of both the CC and the 6C $S(l)$
are shown in Fig.~\ref{fig-pa-cc5c}. Also included in this figure is
an energy dependent empirical potential, Burzynski (6) of
Ref~\cite{pa-edept} evaluated at 40 MeV, which was obtained by
inversion of the complex phase shifts of Burzynski {\em et al},
\cite{burzynski}.  The real, $V_1$ and $V_2$ components for the CC and
6C cases are similar in overall magnitude and shape to the potentials
for the SC case. However the decrease in magnitude of the $V_2$ term
on introducing the coupling does give an improved agreement with the
empirical potential at larger radii.

Significantly, the channel coupling introduces both a generative
imaginary potential at the nuclear centre and a strong parity
dependence in the imaginary potential. This parity dependence, much
larger than that reported in Sect.~\ref{sect-impots}, probably relates
to the influence of the stronger non-locality in the d + $^3$He
channel on the p + $\alpha$ channel.  The imaginary $V_2$ component is
of roughly the magnitude found empirically but the shape shows far
less agreement with the empirical imaginary $V_2$ component than can
be seen for all other components in Fig.~\ref{fig-pa-cc5c}. The
magnitude of the generation is far greater than that found
empirically, but the results of Sect.~\ref{sect-impots} suggest that a
disagreement in the imaginary components is expected at small radii
due to the disagreement of the real components for $r < 2$ fm.  The
agreement for $r > 2$ fm maybe slightly fortuitous, since the
magnitude of the empirical potential depends strongly on the chosen
form of energy dependence, which could not be uniquely established by
inversion, \cite{pa-edept}.  However, the empirical potential shown in
Fig.~\ref{fig-pa-cc5c} does provide the best fit to the phase shifts
of Burzynski {\em et al}.  A consistent energy dependence cannot be
established for the RGM potentials due to the irregularities in the
$S(l)$ in both the CC and 6C cases. However, the potentials displayed
in Fig.~\ref{fig-pa-cc5c} for both cases are accurate to within 0.5
MeV over the energy range 35 -- 45 MeV.

Application of IP inversion to the full RGM calculations, inclusive of
spin-orbit coupling, leads to central parity dependent potentials
which are very close to those shown in Fig.~\ref{fig-pa-cc5c}.  Unlike
the calculations of Sect.~\ref{pa-impots} these
inversions generate a negative imaginary spin-orbit potential of
depth at least $\sim 1$ MeV at maximum, in rough agreement with the
empirical solution, although the empirical imaginary spin-orbit
component is not established very accurately.  A reasonable fit to the
RGM $S(l)$ is possible without a spin-orbit parity dependence, but, by
including the spin-orbit $V_2$ terms in the inversion, $\sigma^2$ is
reduced by a factor of ten. The resulting potentials have a small
radial range and are $< 1$ MeV in magnitude, and again are consistent
with empirical results, \cite{pa-edept}.

\subsection{d + $^3$He}

The changes in the d + $^3$He $S$-matrix arising from introducing the
coupling to the p + $\alpha$ reaction channel, even before the
inclusion of deuteron distortion channels, are much larger than the
equivalent changes found for p + $\alpha$ $S(l)$, \cite{RGM-5n}. This
behaviour is illustrated in Fig.~\ref{fig-sme-dh}. The reaction
channel coupling contributions to the absorption become more
noticeable at the higher of the two energies, predominantly in the d
and f waves, as in the p + $\alpha$ case.  The further addition of the
deuteron pseudo-states produces only small changes in $\arg (S)$, but
these states contribute more significantly to $|S|$, particularly for
$l=0$ due to the existence of a pseudo-resonance at just below 20 MeV.

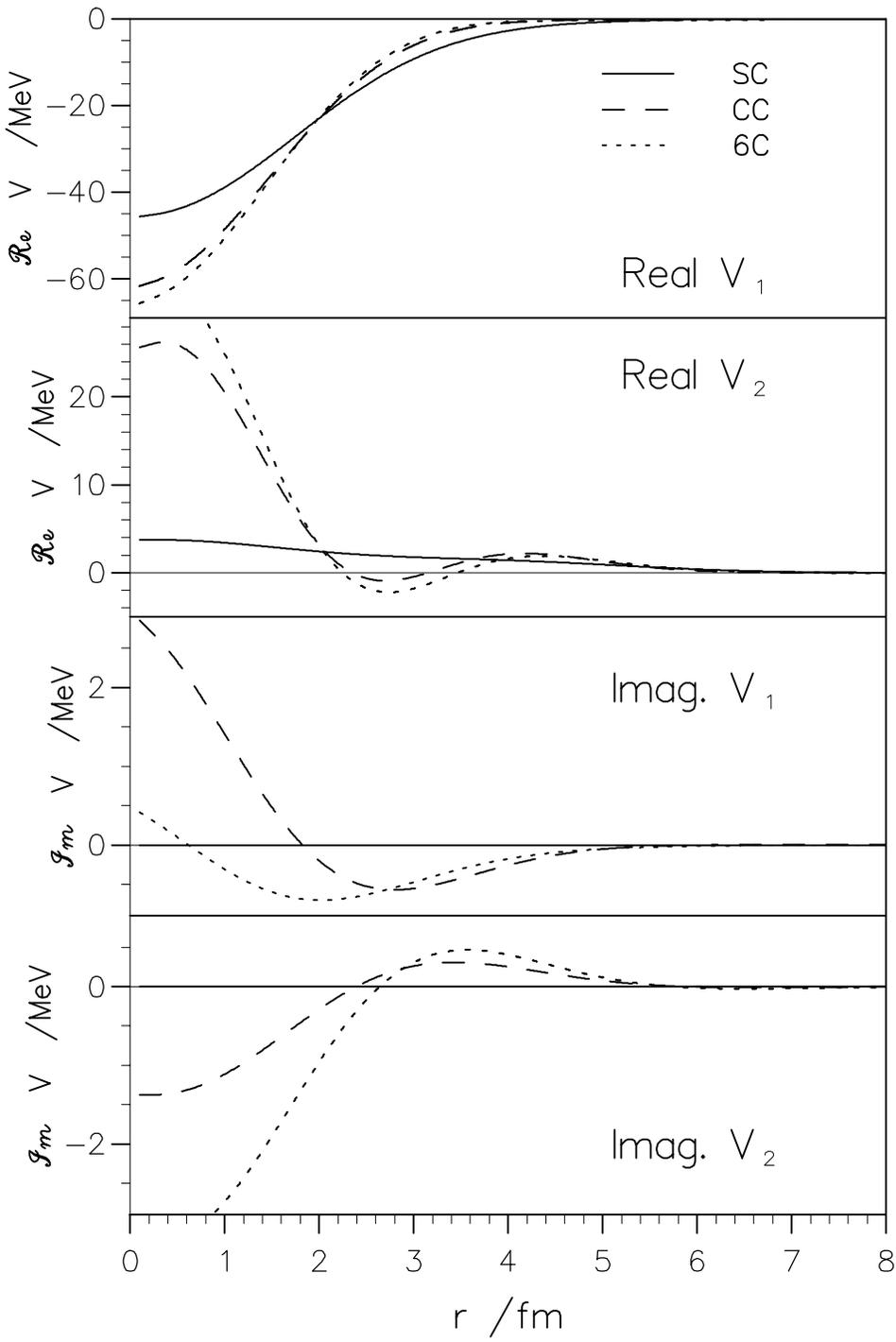
\begin{figure}
\caption{For d + $^3$He at a laboratory energy of 21.67 MeV,
the real and imaginary, $V_1$ and $V_2$ components obtained by inversion of
the CC $S(l)$ (dashed lines) and 6C $S(l)$ (dotted lines) compared
with the equivalent components determined for the SC case (full lines).}
\label{fig-dh-cc5c}
\end{figure}

The potentials obtained by inversion for both the CC and 6C cases at
the laboratory energy of 21.67 MeV are displayed in
Fig.~\ref{fig-dh-cc5c} together with the SC potential.  The
differences noted above for $\arg (S)$ give rise to large differences
between the real CC and SC potentials. The real $V_1$ component for
the CC case is deeper at the centre and has a reduced radial range.
For the even $l$-values, a large positive contribution to $V_2$ at
small radii compensates to a certain extent for the change in the real
$V_1$ term.  The apparent oscillation in $V_2$ probably arises because
the changes in the f-wave introduced with antisymmetry are
effectively nullified by the additional effects of channel coupling.

The imaginary components for the CC case are similar in shape to those
of the p + $\alpha$ CC imaginary potential, reflecting the strong
influence of the channel coupling over the other effects of the
exchange non-locality.  The $V_1$ component is strongly generative at
small radii and the $V_2$ term has a comparable size.  To reproduce
the energy dependence of $|S|$ for this CC case, changes in the shape
of the imaginary $V_2$ term are required.  The large difference for
$|S(l=0)|$ between the CC and 6C cases results in very different
imaginary potentials for these two cases. The precise form of these
potentials is affected by the pseudo-resonance close by in energy and
considerable variations in the imaginary potential are necessary to
describe the energy dependence of $|S|$ around this resonance. However
the most general feature of these calculations is the decrease in the
generation of the $V_1$ component near $r=0$ compared to the CC case.

The full CC calculation, incorporating a spin-orbit interaction, again
leads to central local potentials which are very similar to those
displayed in Fig.~\ref{fig-dh-cc5c}. Accurate inversion of these
$S(l)$, i.e. to obtain a low $\sigma$, requires parity dependent
spin-orbit terms. The resulting imaginary spin-orbit terms are very
small, $\sim 0.2$ MeV at maximum depth, that is little larger than
the equivalent terms reported in Sect.~\ref{scim-dh-s0p5}, so that the
additional non-locality produced by the channel coupling has little
effect on these terms. However, if the imaginary
spin-orbit $V_1$ is excluded from the inversion, the resulting
imaginary $V_1$ central potential becomes oscillatory.

\section{Summary and conclusions}
\label{sect_conc}

RGM calculations followed by $S$-matrix to potential inversion have
been carried out for p + $\alpha$ and d + $^3$He scattering to
investigate the effects of antisymmetry and channel coupling on the
local potential and particularly the imaginary component.  The real
part of the p + $\alpha$ potential has been widely studied in previous
investigations, but the d + $^3$He potential has received considerably
less attention. Here, for probably the first time, d + $^3$He
potentials are established by accurate inversion of phase shifts, for
both channel spins, $s=1/2$ and $s=3/2$ and a study is presented of
the changes in the potentials due to deuteron distortion and channel
coupling to the p + $\alpha$ configuration (for $s=1/2$).

The real parts of the p + $\alpha$ potential are essentially well
determined by the direct contribution plus exchange terms and these
terms are only slightly modified by the presence of absorption,
whether introduced through channel coupling or via a phenomenological
potential. Consequently the strong agreement between the real
empirical potential and potentials local equivalent to the single
channel RGM established in Ref.~\cite{pa-edept}, up to 65 MeV, is
unsurprising. However the inclusion of a single reaction channel does
improve the agreement of the empirical and RGM potentials at the
lab. energy of 40 MeV at least, probably to the limit to which the
empirical potential is established.

The shape of the imaginary potential is determined by both the
exchange non-locality and the particular reaction channels open.  Even
the simplest form of absorption, for example that represented by
introducing a simple imaginary term into the single channel RGM, is
significantly distorted by the non-locality arising through exchange
alone.  For p + $\alpha$, this effect leads to a reduction in the
absorption for low $l$-values and may lead to an emissive potential at
the nuclear centre. This behaviour is associated with the Perey
effect, which is characteristic of one nucleon exchange.  The more
complicated exchange non-locality of antisymmetrisation
may also introduce parity dependence into the imaginary
components.  The effect is not significant for p + $\alpha$, but does 
generate a strong imaginary parity dependence for d + $^3$He.  This
parity dependence suggests a considerable influence of the heavy-particle
pickup contribution in deuteron scattering which appears related to
the large radial extent of the deuteron.  A fully antisymmetrised
calculation inclusive of reaction channels also leads to a strong
imaginary parity dependence for both p + $\alpha$ and d + $^3$He,
although the link to specific exchange effects is more difficult
to assign with reaction channel coupling.

The imaginary potential components determined from the reaction channel RGM
calculations have roughly the correct magnitude and are qualitatively
similar in shape to the empirical imaginary potentials. The imaginary
parity dependence is the least accurate component, but the
disagreement is not surprising due the underprediction of these RGM
calculations for the absorption of certain partial waves.  Deuteron
distortion contributions have little effects on the p + $\alpha$
potential and inclusion of additional reaction channels does not
correct the underestimation of the d-wave absorption, \cite{KKT-95}.
However, a critical omission in these RGM calculations may be the lack of a
tensor component which would allow coupling to the $s=3/2$ d + $^3$He
channel.  The inclusion of spin-orbit coupling in the RGM introduces
additional components in the local potential, that of most
significance being an imaginary spin-orbit term. This component roughly
agrees with  the imaginary spin-orbit term found in the
empirical potential

The description of d + $^3$He is more complicated due to the high
compressibility of the deuteron. The real potentials for both values
of channel spin show the strong increase in magnitude on introduction
of exchange effects, as found for p + $\alpha$ and in many other few
nucleon systems. This behaviour is commonly associated with knock-on
exchange. The real local potential is also strongly parity dependent,
although the precise shape of this parity dependence differs quite
significantly for the two channel spins as expected from the phase
shift behaviour.  The imaginary terms have a similar large dependence
on the channel spin, but the resultant shape of the calculated
potentials must necessary be qualitative due to the strong dependence
on the number of reaction channels included.

The potentials determined in this study differ considerably in shape
from the Saxon-woods form used in many standard optical analyses, and
expose the limitations of such analyses.  For the cases considered
here, the imaginary central component and both real and imaginary
parity dependent forms in particular need to be described by a
flexible parameterisation. It is possible that similar properties may
be found for other light nucleon systems.  The application of
inversion to p + $^{16}$O elastic cross-section and analysing power
data not only lead to a far better reproduction of the data than could
be achieved with conventional optical analyses, but also established
imaginary and parity dependent potentials qualitatively similar to
those presented here. 

This study suggests a need to investigate how far the contributions of a
fully antisymmetrised coupled reaction channel analysis can account
for the empirical p + $^{16}$O potential, in particular the shape of
the imaginary components which show a strong similarity to the
potentials found in this work. Also of interest is a empirical
analysis of d + $^3$He, but this presents a much harder proposition.
Independent of any tensor components, different parameterisations are
necessary for the two channel spins and parity dependence is expected
in both cases. A very comprehensive data set is then necessary to
establish all these components to a reasonable accuracy.

\section*{Acknowledgements}
The author is very grateful to Dr. R.S. Mackintosh for useful
discussions and for a careful reading of this manuscript.  Financial
support is acknowledged from the EPSRC of the UK, under the grant
GR/H00895.

\setlength{\parindent}{0.0 in}

\newpage

\begin{figure}
\caption{For p + $\alpha$, values of $|S_l|$ and $\arg(S_l)$ calculated
at energies of 30 and 40 MeV for the following cases (defined in
Sect.~\ref{rgm-defs}), DIR (full lines), IM (dashed
lines),  CC (dotted lines) and 6C (dash-dot lines).}
\label{fig-sme-pa}
\end{figure}

\vfill


\clearpage

\newpage

\begin{thebibliography}{99}
\bibitem{pa-emp} S.G. Cooper and R.S. Mackintosh, Phys. Rev.  C{\bf 43},
 (1991) 1001 .
\bibitem{pa-edept} S.G. Cooper and R.S. Mackintosh, Phys. Rev.  C{\bf 54}
 (1996) 3133.
\bibitem{p-ox16} S.G. Cooper, Nucl. Phys. {\bf A618} (1997) 87.
\bibitem{na-exch} D. R. Thompson and Y.C. Tang,  Phys. Rev. {\bf C4}
(1971) 306.
\bibitem{pa-rgm} S.G. Cooper, R.S. Mackintosh, A. Cs\'ot\'o, and R.G.
Lovas, Phys. Rev.C {\bf 50} (1994) 1308.
\bibitem{n-nucl} S.G. Cooper and R.S. Mackintosh, Nucl. Phys. {\bf A592}, 
 (1995) 338
\bibitem{Reichstein}  I. Reichstein and Y.C. Tang, Nucl. Phys. {\bf A158},   
529 (1970).
\bibitem{RGM-dhe} F.S. Chwieroth, Ronald E. Brown, Y.C. Tang and D.R. 
Thompson, Phys. Rev. {\bf C8} (1973) 938.
\bibitem{RGM-5n} F.S. Chwieroth, Y.C. Tang and D.R. Thompson, Phys. Rev. 
{\bf C9} (1973) 56.
\bibitem{RGM-5n-im} F.S. Chwieroth, Y.C. Tang and D.R. Thompson, Phys. Lett.
{\bf 46B} (1973) 301.
\bibitem{RGM-5n-d} H. Kanada, T. Kaneko and Y.C. Tang, Nucl. Phys. {\bf A504}
(1989) 529.
\bibitem{KKT-95} H. Kanada, T. Kaneko and Y.C. Tang, Prog. Theor. Phys.
{\bf 94} (1995) 1061.
\bibitem{pa-res} G. Bl\"{u}ge, and  K. Langanke, Phys. Rev. {\bf C41}
(1990) 1191. \\
-- A. Cs\'{o}t\'{o}, R. G. Lovas and A. T. Kruppa, Phys. Rev. Lett.
{\bf 70} (1993) 1389.
\bibitem{RGM-code} G. Bl\"{u}ge, K. Langanke and H.-G. Reusch, ``Computational
Nuclear Physics 2'', Ed. K. Langanke, J.A. Maruhn and S.E. Koonin (1993)
(Springer-Verlag, New York).
\bibitem{ad-dist} H. Kanada, T. Kaneko, S. Saito and Y.C. Tang, Nucl. Phys.
{\bf A444} (1985) 209.
\bibitem{A-selove} F. Ajzenberg-Selove, Nucl. Phys. {\bf A490} (1988) 209.
\bibitem{kuk-Ryz} V.I.Kukulin and G.G. Ryzhikh, Prog. Part. Nucl. Phys.
{\bf 34} (1995) 397.\\
-- V.I.Kukulin and V.N. Pomerantsev, Sov. J. Nucl. Phys. {\bf 50} (1989) 17.
\bibitem{kuk-priv} V.I.Kukulin, private communication.
\bibitem{minn-pot} D. R. Thompson, M. LeMere and Y.C. Tang, Nucl. Phys.
{\bf A286} (1977) 53.
\bibitem{LeMere} M. LeMere, Y. Fujiwara, Y.C. Tang and Q.K.K. Liu, Phys. Rev. 
C{\bf  26},  1847 (1982).
\bibitem{plattner} G.R. Plattner, A.D. Bacher, and H.E. Conzett,  Phys. Rev. 
C{\bf  5},  1158 (1972).
\bibitem{burzynski} S. Burzynski, J. Campbell, M. Hammans, R. Henneck,
W. Lorenzon, M.A. Pickar, and I. Sick,  Phys. Rev. C{\bf 39}, 56 (1989).
\bibitem{imago}S.G. Cooper and R.S. Mackintosh, Users manual for IMAGO,
Open University preprint OUPD9201.
\bibitem{edep-na} R.S. Mackintosh and S.G. Cooper, J. Phys. G:Nucl. Part. 
Phys.{\bf 23} (1997) 565.
\bibitem{n-wavefns} S.G. Cooper and R.S. Mackintosh,, Nucl. Phys. {\bf A511} 
(1990) 29
\bibitem{RGM-dist} H. Kanada, T. Kaneko, P. N. Shen and Y.C. Tang, 
Nucl. Phys. {\bf A457} (1986) 93.
\end{thebibliography}
\end{document}